# Nonreciprocal Yagi-Uda Filtering Antennas

J. W. Zang, X. T. Wang, A. Alvarez-Melcon, *Senior Member, IEEE*, and J. S. Gomez-Diaz, *Senior Member, IEEE*

*Abstract*— This paper proposes a novel and compact nonreciprocal filtering antenna based on temporal modulation. The device is composed of a third order filtering section integrated into a Yagi-Uda planar printed antenna. Strong nonreciprocity in transmission and reception is achieved at the same operation frequency by time-modulating the resonators of the filtering section. These resonators are implemented as quarter wavelength microstrip lines terminated with varactors located on the ground plane. Such plane is also employed as a reflector for the Yagi-Uda antenna and to host the coplanar waveguides that feed low-frequency signals to the varactors, thus leading to a very compact design. A prototype is manufactured and successfully tested at 2.4 GHz, showing isolations greater than 20 dB between transmission and reception modes in both E- and H-planes for all directions in space and a gain drop of only 3.5 dB compared to a reference antenna. The proposed devices can easily be integrated with other electronic circuits and may find exciting applications in communication, radar, and sensing systems.

*Index Terms*— Filtering antennas, nonrecipocal devices, spatio-temporal modulations, time modulated resonators.

## I. INTRODUCTION

Filters and antennas are crucial components to wireless communication systems. Traditionally, these components are designed separately and then connected using transmission lines. Using this standard approach, the problems of relatively large volume and degraded frequency selectivity are unavoidable due to the extra losses and reflections introduced by the extra lines. Recently, the concept of filtering antennas has been put forward [1-4]. These structures integrate a bandpass filter and an antenna into a unique structure without requiring additional matching circuits. This approach leads to compact and low-profile devices that exhibit good radiating and selective responses together with reduced insertion losses.

Several types of filtering antennas have been studied in the past, including monopole [1]-[4], patch [5]-[8], and slot [9]-[11] configurations. In addition, low-profile metasurface-based filtering antennas have led to relatively high gain performances [12], [13]. Moreover, filtering functionalities through dielectric resonator antennas have also been applied to enhance the operational bandwidth [14], [15]. The performance of filtering antennas can be further enhanced by loading them with tunable components such as varactors or pin diodes, which lead to pattern [16] and/or frequency [17] reconfigurable capabilities.

Manuscript received Month x, 2019.

This work is supported in part by the National Science Foundation with CAREER Grant No. ECCS-1749177 and by the grants PRX18/00092 and TEC2016-75934-C4-4-R of MECD, Spain.

J. W. Zang and X. T. Wang are with the School of Information and Electronics, Beijing Institute of Technology, Beijing 100081, China. J. W. Zang was a visiting student in the Department of Electrical and Computer Engineering, University of California at Davis, Davis, CA 95616 USA.

A. A. Melcon was on sabbatical in the Department of Electrical and Computer Engineering at UC Davis. He is now with the Department of Information and Communication Technologies, Technical University Cartagena, 30202, Spain.

J. S. Gomez-Diaz is with the Department of Electrical and Computer Engineering, University of California at Davis, Davis, CA 95616 USA. (e-mail: jsgomez@ucdavis.edu).

In a related context, nonreciprocal electromagnetic wave propagation is needed in many practical applications. For example, isolators are commonly used to protect a transmitter from its strong reflection due to impedance mismatch at the antenna terminals. Also, circulators are essential components to enable the transmitter and the receiver to share a common antenna. However, nonreciprocal components have historically relied on ferrites that require a magnetic bias to break time-reversal symmetry. Magnetic-based nonreciprocal components are incompatible with integrated circuit (IC) technology, and they tend to be bulky and lossy [18]. To overcome these challenges, spatiotemporal modulations approaches have been proposed to break reciprocity without relying on magnetic materials. This technique has been found useful to realize a wide variety of efficient, compact, and IC-compatible nonreciprocal components, including isolators [19]-[22], circulators [23]-[26], and leaky wave antennas [21], [27], [28]. Besides, time-modulated metasurfaces can provide exotic functionalities such as nonreciprocal beam scanning and/or focusing [29] and serrodyne frequency translation [30]. Very recently, nonreciprocal microstrip filters based on time modulated resonators have been proposed and experimentally demonstrated [31, 32]. Several prototypes based on both lumped elements and on distributed resonators were fabricated and successfully measured, showing isolation levels higher than 15 dB over 2% fractional bandwidths.

In this Letter, we propose, analyze, design, fabricate, and characterize a nonreciprocal planar filtering antenna that exploits time-modulated resonators to break reciprocity between transmission and reception operation at the same operational frequency. The device integrates a nonreciprocal filter with a planar printed Yagi-Uda antenna. The proposed magnetic-free antenna efficiently radiates power in transmission while it rejects power from all directions during reception. In addition to non-reciprocity, the radiation patterns also exhibit good selectivity and abrupt behavior with frequency, typical of traditional filtering antennas. Compared to other nonreciprocal antenna configurations previously studied in the literature [21], [27]-[29] the proposed device (i) radiates and receives energy at the same frequency without undergoing any frequency conversion; and (ii) is completely angle-insensitive. Measured results show radiation patterns with a main beam directed towards endfire, with a high isolation between transmission and reception modes that is greater than 20 dB at 2.4 GHz and a drop of the gain of just 3.5 dB as compared to a reference antenna.

## II. DESIGN OF NONRECIPROCAL FILTERING ANTENNAS

The proposed device is illustrated in Fig. 1. The filtering capability is mainly obtained from three coupled resonators that are implemented with quarter wavelength microstrip lines and terminated on one side with a varactor. They directly feed a printed Yagi-Uda antenna composed of an active dipole element and two passive directors. To break time-reversal symmetry, a sinusoidal modulating signal of frequency $f_m$ is applied to the different varactors. This modulation signal imparts different phases ($\varphi_1, \varphi_2, \varphi_3$) to the resonators and transform the filtering antenna into a non-linear component [31,32]. Strong nonreciprocity appears due to nonreciprocal couplings and phase cancellation effects among the nonlinear harmonics that are generated in the time-modulated resonators, as detailed in [32]. In the following, we demonstrate that it is possible to realize significantly different transmission and reception of RF signals by properly optimizing the frequency, amplitude, and phase of the modulation signals.

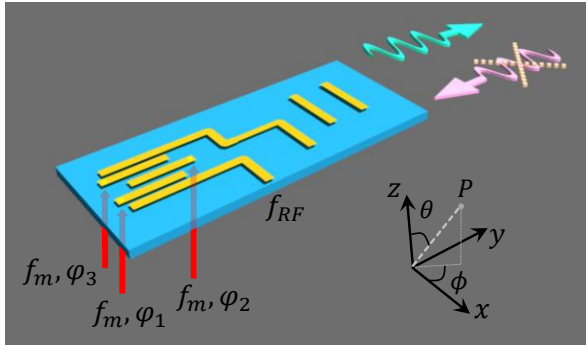

Fig. 1. Proposed nonreciprocal Yagi-Uda filtering antenna. The resonators that compose the filtering part are time-modulated with identical frequency $f_m$ but with different phases. The antenna radiates towards endfire ($\theta = 90°$) at $f_{RF}$ but cannot receive energy at the same frequency.

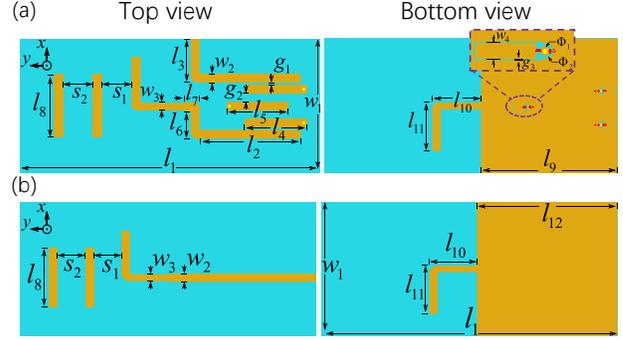

Fig. 2: Proposed nonreciprocal Yagi-Uda antenna (top row) and a reference antenna employed for comparison purposes (bottom row). The first and second columns show the top and bottom views of the devices printed on a Rogers DiClad 880 substrate with a thickness of 1.575 mm, a relatively dielectric constant of 2.2, and a loss tangent of 0.0009. The dimensions of the devices are: $l_1$ = 170 mm, $l_2$ = 53.9 mm, $l_3$ = 28.2 mm, $l_4$ = 32.4 mm, $l_5$ = 14.2 mm, $l_6$ = 8.4 mm, $l_7$ = 34 mm, $l_8$ = 80 mm, $l_9$ = 28.5 mm, $l_{10}$ = 29 mm, $w_1$ = 80 mm, $w_2$ = 4.84 mm, $w_3$ = 4 mm, $w_4$ = 2.6 mm, $s_1 = s_2$ = 16 mm, $g_1$ = 0.36 mm, $g_2$ = 4.16 mm, $g_3$ = 0.2 mm, $\Phi_1$ = 1 mm, $\Phi_2$ = 2 mm.

The non-reciprocal Yagi-Uda filtering antenna is printed on the two sides of a dielectric board, as shown in Fig. 2 (a). A half ground plane is printed on the bottom side of the dielectric board to (i) act as a passive reflector for the Yagi-Uda antenna; and (ii) host the microstrip line resonators printed on the top side (see Fig 2). Within this setup, the antenna provides endfire radiation patterns with the main beam pointing towards $\theta$ = 90°, as illustrated in Fig. 1. The RF signal ($f_0$) is fed to the input port of the filtering structure using a microstrip line with 50 Ω characteristic impedance. The output port of the filtering section directly drives the active dipole of the Yagi-Uda antenna. The key element to induce non-reciprocity in this device is to time modulate the resonators of the filtering structure with a sinusoidal signal of frequency $f_m$. To this purpose, the quarter wavelength resonators are terminated on one end with a plated hole (0.2 mm in radius) connected to a varactor [details are shown in the inset of Fig. 2(b)]. Short coplanar waveguides (CPW) are etched on the common ground plane and are used to feed the modulation signals into the varactors. An inductor of value 180 nH is used as choke to increase the isolation between RF ($f_0$) and modulation signals ($f_m$). Measured data confirms that such isolation is greater than 30 dB. SMV1231 varactors from Skyworks Solutions are employed and set to operate in reverse bias state. Their static capacitance is varied with time according to the modulation signals as

$$C_i(t) = C_0[1 + \Delta_m cos(\omega_m t + \varphi_i)] \quad (1)$$

where $\omega_m$ is the angular frequency of the modulation signal ($\omega_m = 2\pi f_m$), $C_0$ is the static capacitance, and $\Delta_m = \frac{\Delta_C}{C_0}$ is the modulation index ($0 < \Delta_m < 1$), with $\Delta_C$ being the maximum capacitance variation introduced by the modulation signal. It can be observed that all varactors are modulated with signals of same amplitude and frequency. However, a progressive phase shift ($\Delta_\varphi$) is applied between consecutive resonators, setting $\varphi_i = (i - 1)\Delta_\varphi$, with $i = 1,2,3$. Additional details to analyze and design nonreciprocal filters can be found in 32]. To assess the performance of the proposed nonreciprocal antenna, its response will be compared against the reference Yagi-Uda antenna [33] described in Fig. 2(b). This device exhibits identical radiating part as the original filtering antenna.

To design the proposed nonreciprocal devices, we first independently consider its radiation [34] and filtering parts [31], [32]. For the latter, we assume that the nonreciprocal filter is loaded with the input impedance of the Yagi-Uda antenna. Besides from this precaution, the design of the nonreciprocal filter follows standard techniques to adjust the coupling elements ($g_1$ and $g_2$ in Fig. 2) and the length of the resonators ($l_4, l_5$ in Fig. 2), aiming to obtain an equi-ripple response inside the passband [5]. The filter is designed for the following target specifications: center frequency $f_0$ = 2.4 GHz, fractional bandwidth $FBW = 4\%$ and return losses $RL$ = 13 dB. The radiating part is designed following standard procedures to realize Yagi-Uda configurations [34].

Once the design of the radiating and filtering parts is completed, the two configurations are connected to compose a single device. A final optimization is then carried out to compensate for cross couplings and loading effects between the different elements of the structure. This final optimization step gives the dimensions shown in the caption of Fig. 2. Both designs have been manufactured using a Rogers DiClad 880 substrate, with a thickness of 1.575 mm, relatively dielectric constant of 2.2, and loss tangent of 0.0009. Photographs of the fabricated nonreciprocal Yagi-Uda filtering antenna and the reference antenna are given in the insets of Fig. 3. In the inset of the panel (a) we can clearly identify the three connectors used to introduce the modulating signals ($f_m$) that are located in the bottom ground plane and the connector employed to feed the RF signal ($f_0$) into the device.

It is important to emphasize that, up to now, the design techniques that we have employed involve classical methods for reciprocal antennas and filters. The design of the components at this stage was carried out with the full-wave simulation software tool Ansys HFSS [36]. We remark that we have not considered yet the presence of the modulation signals that will time-modulate the resonators. As detailed in the following, the filtering part of the device will become nonreciprocal by properly adjusting the relevant parameters of the modulating signals, namely the frequency ($f_m$), the static capacitance ($C_0$), the modulation index ($\Delta_m$) and the phase shift between resonators ($\Delta_\varphi$) [31], [32].

### III. ANTENNA PERFORMANCE

The magnitude of the device reflection coefficient is shown in Fig. 3a. First, we have tested the device without applying time-modulation signals (static case). In the experiments, the varactors are biased with a DC bias to obtain an equi-ripple response within the passband. The adjustment of the DC voltages permits to compensate for manufacturing errors in the resonant frequencies of the resonators. However, potential errors introduced in the coupling coefficients cannot be compensated. This is the reason why the measured response (red line) shows slightly narrower bandwidth than numerical simulations (blue line). The 10-dB fractional bandwidth is 4.4% (2.333-2.437 GHz) and 2.6% (2.354-2.417 GHz) for simulation and measurement, respectively.

As discussed above, time-reversal symmetry can be broken in this device by applying time-modulation to the resonators of the filtering part. Strong nonreciprocity between transmission and reception operation can be obtained by properly optimizing the parameters of the modulating signals. After following the procedure detailed in [31], [32], we find the following quasi-optimized values for our proposed configuration: $f_m$=75 MHz, $C_0$=0.86 pF, $\Delta_m$=0.09, and $\Delta_\varphi$=56°. It can be observed that the design leads to weak modulation signals (small modulation frequency and index) that are common in other implementations [28], [29] and permit to operate the varactors in their linear regime. The measured reflection coefficient in the presence of modulation signals is shown with black solid line in Fig. 3a. The antenna tends to have a slightly narrower bandwidth as compared to the static situation, although it still exhibits good frequency selectivity. The 10-dB fractional bandwidth is of 1.7 % (2.376-2.416 GHz). Fig. 3b shows the reflection coefficient of the reference Yagi-Uda antenna, which exhibits good matching at 2.4 GHz and poor selectivity in frequency.

We next investigate the radiation properties of the proposed filtering antenna. For the sake of simplicity, in this study we normalize all radiation patterns with respect to the maximum value given by the reference antenna. The simulated and measured radiation patterns of the reference antenna in E-and H-planes at 2.4 GHz are shown in Fig. 4a. It can be observed very good agreement between numerical simulations and measured data. The radiation patterns show a main beam with 6 dB of directivity, typical of these electrically small antennas [34]. The radiation patterns of the proposed filtering antenna without applying time-modulated signal (i.e., the static case) are shown in Fig 4(b). Again, good agreement is observed between simulation and measurements. We remark that the operation frequency is inside the passband of the integrated filter and thus the radiation patterns are very similar to those obtained for the reference antenna. Fig. 4(c) shows the measured radiation patterns after applying the time-modulation signals, distinguishing between transmission (solid blue) and reception (dashed red) modes. It can be observed that the proposed Yagi-Uda filtering antenna can efficiently radiate into space and that there is only 3.5 dB gain loss compared to the reference antenna. Strikingly, the received power is about 20 dB lower compared to the level of the transmitted power for both E- and H- planes in all directions of space. As expected, the radiation patterns in transmit mode show again very similar behavior as compared to the reference antenna. In particular, the patterns still exhibit a main beam with 6 dB of directivity towards endfire. This demonstrates that the extra circuits needed to feed the modulating signals have small impact on the radiation properties of the antenna, and thus, clean endfire radiation patterns can still be obtained.

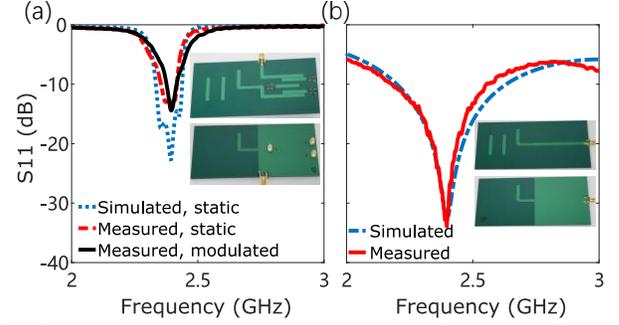

Fig. 3: Measured and numerically simulated reflection coefficient (magnitude in dB) of (a) the proposed nonreciprocal Yagi-Uda antenna detailed in Fig. 2a; and (b) the reference Yagi-Uda antenna described in Fig. 2b. Insets show pictures of the fabricated devices.

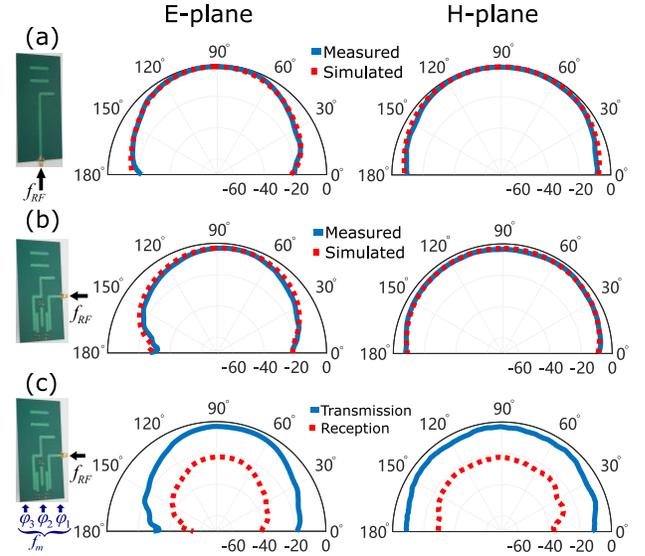

Fig. 4: Radiation patterns (in dB) generated at 2.4 GHz by the devices described in Fig. 2 in the E- (left column) and H- (right column) planes. (a) Reference Yagi-Uda antenna detailed in Fig. 2b. (b) Proposed nonreciprocal Yagi-Uda filtering antenna detailed in Fig. 2a without applying time-modulated signals (static case). (c) Similar to (b) but applying the time-modulation signals to the circuit (modulated case). Results are normalized with respect to the maximum gain provided by the reference antenna.

As a final test, Fig. 5 (a), (b) and (c) show the measured boresight directivity response versus frequency for the reference antenna and for the static and modulated states of the proposed nonreciprocal Yagi-Uda antenna, respectively. For the sake of comparison, all results shown in Fig. 5 are normalized with respect to the maximum value given by the reference antenna in Fig 5 (a). To check nonreciprocity, both the transmission and reception performances are measured. Fig. 5(a) confirms that the reference Yagi-Uda antenna is reciprocal and that it exhibits identical transmission and reception properties. As expected, the antenna directivity exhibits poor selectivity versus frequency. Fig. 5(b) shows that the proposed antenna is reciprocal when time-modulations signals are not applied. Compared to the reference antenna, the static filtering antenna exhibits an improved frequency selectivity (with an out-of-band rejection level over 11 dB for the entire out-of-band and a 3-dB fractional bandwidth of 3.2%) and a drop in the maximum gain of 3 dB due to the extra losses introduced by the filtering section, the varactors, and related biasing components.

Finally, Fig. 5(c) presents the response of the proposed filtering antenna after the time-modulating signals are fed into the device. Strong isolation between transmission and reception modes greater than 20 dB appears at the operation frequency of 2.4 GHz. The modulated device exhibits a similar frequency selective response as in the static case together with an isolation larger than 5dB over the -3dB fractional bandwidth (2.357-2.429 GHz). The maximum gain is 3.5 dB below the reference antenna and 0.5 dB lower than in the static mode. We attribute the extra 0.5 dB of loss to the power that has been converted from the fundamental frequency to higher order non-linear harmonics and has not been converted back to the fundamental frequency [31], [32]. We stress again that, in contrast to the approaches previously presented in the literature that achieve nonreciprocity between the fundamental frequency and a given nonlinear harmonic [21], [27]-[29], our proposed design provides strong isolation between transmission and reception modes at the same operation frequency ($f_0$) without undergoing any frequency conversion.

## IV. CONCLUSION

We have demonstrated a compact and magnetless nonreciprocal planar filtering antenna that exhibits large isolation between transmission and reception modes. The proposed device is composed of a third order filtering section integrated with a planar Yagi-Uda antenna. Strong nonreciprocity is achieved by time modulating the resonators of the filtering section. Compared to other nonreciprocal antennas available in the literature, the filter antenna proposed here does not rely on frequency conversion and thus exhibits nonreciprocity at the same operation frequency. Measured data from a fabricated prototype demonstrated isolation levels greater than 20 dB at the design frequency of 2.4 GHz for both E- and H- planes and all directions in space, with an overall loss of just 3.5 dB compared to a reciprocal reference antenna. We envision that this technology will open new and exciting venues in sensing, communication, and radar systems as well as in security and defense applications.

ACKNOWLEDGMENT

Authors wish to thank Rogers Corporation for the generous donation of the dielectrics employed in this work. We also thank Prof. X. Liu and Prof. R. Branner (University of California, Davis) for providing access to the equipment and anechoic chamber room required to carry out this work.

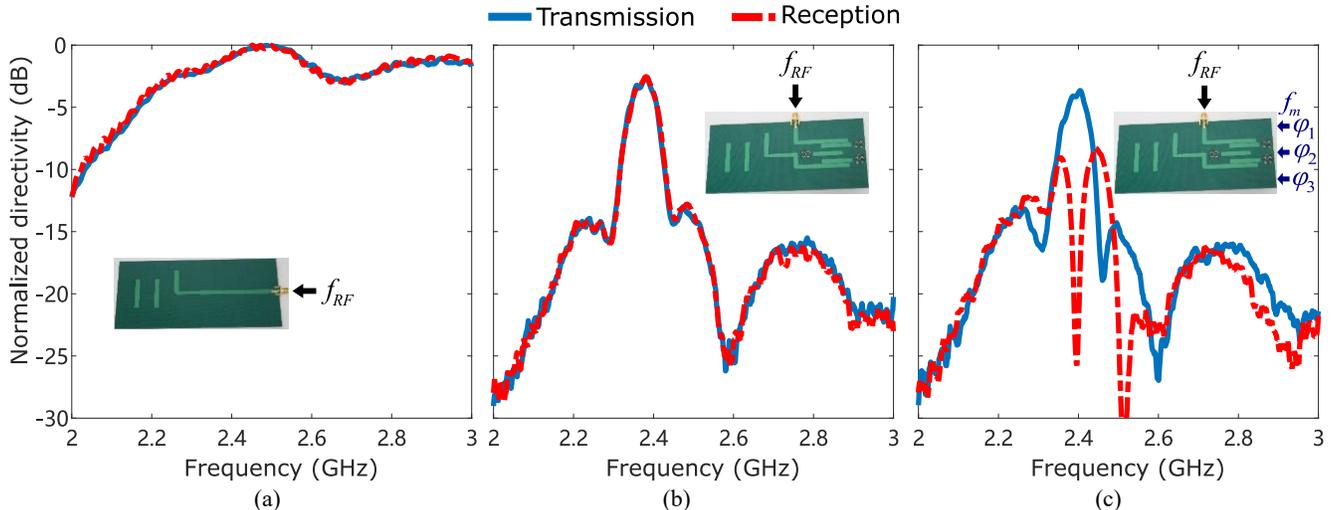

Fig. 5: Normalized directivity of the antennas described in Fig. 2 operating in transmission (solid blue line) and in reception (solid red line) modes. (a) Reference Yagi-Uda antenna. (b) Static state (without applying time-modulated signal) of the proposed nonreciprocal Yagi-Uda filtering antenna. (c) Modulated state of the proposed nonreciprocal Yagi-Uda filtering antenna. Results, normalized with respect to the maximum gain of the reference antenna, are computed versus frequency at the boresight direction.